\documentclass[conference]{IEEEtran}

\usepackage{amssymb}
\usepackage{graphicx}
\usepackage{epstopdf}
\usepackage{gensymb}
\usepackage{color}
\usepackage{amsmath}
\usepackage{bm}
\usepackage{etoolbox}
\usepackage{cite}
\usepackage{array}
\usepackage{booktabs}
\usepackage{multirow}
\usepackage{comment}

\usepackage{threeparttable} 
\usepackage{mathtools}
\usepackage{pifont}

\IEEEoverridecommandlockouts
\IEEEpubid{\makebox[\columnwidth]{\copyright~2019 IEEE. Permission from IEEE must be obtained except personal usage.} \hspace{\columnsep}\makebox[\columnwidth]{ }}

\DeclarePairedDelimiter\ceil{\lceil}{\rceil}

\begin{document}
\title{Building Secure SRAM PUF Key Generators on Resource Constrained Devices}

\author{
   \IEEEauthorblockN{Yansong Gao\IEEEauthorrefmark{1}\IEEEauthorrefmark{2}, Yang Su\IEEEauthorrefmark{3}, Wei Yang\IEEEauthorrefmark{1}, Shiping Chen\IEEEauthorrefmark{2}, Surya Nepal\IEEEauthorrefmark{2}, and Damith C.~Ranasinghe\IEEEauthorrefmark{3}}
  \IEEEauthorblockA{\textit{\IEEEauthorrefmark{1}School of Computer Science and Engineering, Nanjing University of Science and Technology}, China}
   \IEEEauthorblockA{\textit{\IEEEauthorrefmark{2}Data61, CSIRO}, Syndey, Australia}
   \IEEEauthorblockA{\textit{\IEEEauthorrefmark{3}Auto-ID lab, School of Computer Science, The University of Adelaide}, Adelaide, Australia
  \\ yansong.gao@njust.edu.cn;  yang.su01@adelaide.edu.au; generalyzy@gmail.com; shiping.chen@data61.csiro.au; \\ surya.nepal@data61.csiro.au; damith.ranasinghe@adelaide.edu.au}\vspace{-1.0cm}
}

\maketitle

\begin{abstract}
A securely maintained key is the premise upon which data stored and transmitted by ubiquitously deployed resource limited devices, such as those in the Internet of Things (IoT), 
are protected. However, many of these devices lack a secure non-volatile memory (NVM) for storing keys because of cost constraints. Silicon physical unclonable functions (PUFs) offering unique device specific secrets to electronic commodities are a low-cost alternative to secure NVM. 
As a physical hardware security primitive, reliability of a PUF is affected by thermal noise and changes in environmental conditions; consequently, PUF responses cannot be directly employed as cryptographic keys. A fuzzy extractor can turn noisy PUF responses into usable cryptographic keys. However, a fuzzy extractor is not immediately mountable on (highly) resource constrained devices due to its implementation overhead. We present a methodology for constructing a lightweight and secure PUF key generator for resource limited devices. In particular, we focus on PUFs constructed from pervasively embedded SRAM in modern microcontroller units and use a batteryless computational radio frequency identification (CRFID) device as a representative resource constrained IoT device in a case study. 
\end{abstract}


\IEEEpeerreviewmaketitle

\section{Introduction}
Resource limited IoT (Internet of Things) devices provide challenging environments for building privacy and security preserving mechanisms. 
Although various cryptographic algorithms can be engineered to address the aforementioned challenges, all of these measures eventually rely on a securely maintained key. Nowadays, digital keys are stored in non-volatile memory (NVM) such as FLASH that is often assigned externally to a computing platform. However, it has been shown that securing digital key storage is non-trivial in practice due to e.g., technical or cost limitations~\cite{hiller2016key}.
Silicon physical unclonable functions (PUFs)~\cite{suh2007physical,roel2012physically}  provide an alternative. PUFs can replace the functionality of the NVM based keys while increasing the security level of cryptographic key storage using standard CMOS technology based hardware security primitives. 
Overall, silicon PUFs offer: i) low fabrication costs~\cite{hiller2016key}; ii) uniqueness (an inseparable \textit{fingerprint} of a hardware instance); and iii) enhanced resistance to attacks, especially, invasive attacks~\cite{tuyls2006read,gassend2008controlled,obermaier2018measurement}.



\begin{figure}[b!]
	\centering
	\includegraphics[trim=0 0 0 0,clip,width=0.45\textwidth]{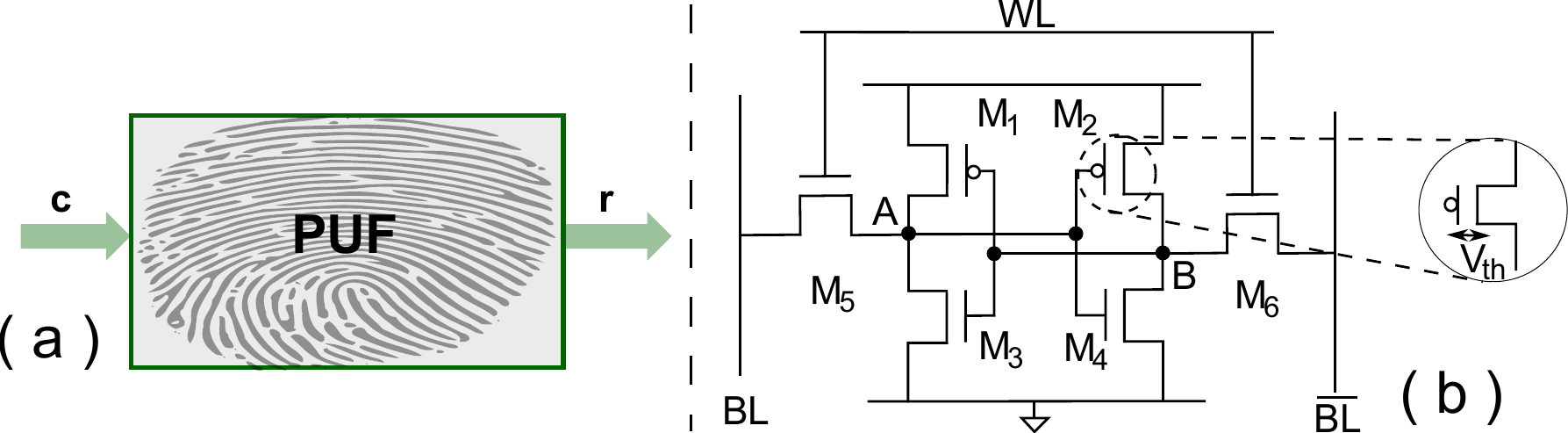}
	\caption{(a) Queried by a challenge, the PUF produces an instance-specific response. (b) SRAM PUF: Threshold voltage  $V_{\rm th}$ mismatches of the transistors determines the response~\cite{holcomb2009power}. For example, when the $ V_{\rm th, M_1} $ is slightly smaller than $ V_{\rm th, M_2} $, at power-up, the transistor M$_1 $ starts conducting before M$_2 $, thus, $ {\rm A}=$ `1'. This in turn prevents M$_2 $ switching on. As a consequence, the SRAM power-up state prefers to be `1' ($ {\rm A}= $ `1', $ {\rm B}= $`0'), which is the response, while the address is the challenge. }
	\label{fig:SRAMPUF}
\end{figure}
\vspace{1mm}

\noindent\textbf{Physical Unclonable Functions (PUFs): }
A PUF, in essence, extracts secrets from inevitable process variations that is alike device fingerprints~\cite{suh2007physical,chang2017retrospective,gao2017puf}. Hence, in reality, identical PUF instances cannot be forged, not even by the same manufacturer. As illustrated in Fig.~\ref{fig:SRAMPUF}(a), a PUF can be abstracted as a function where a given binary input (challenge $\bf c$) applied to different PUF instances produces differing instance-specific binary outputs (responses $\bf r$). 
According to the number of challenge response pairs (CRPs) yielded, a PUF is generally categorized into strong and weak PUF classes~\cite{herder2014physical}. Generally, a strong PUF yields CRPs exponential with respect to the size of the challenge or area. Conversely, its counterpart, weak PUF, yields limited number of CRPs---CRP space linearly increases with the area or the size of the challenge. Elementary PUF applications are: i) lightweight authentication; and ii) key generation. 

\noindent\textbf{Lightweight Authentication: } PUFs provide novel avenues for realizing lightweight authentication mechanisms, usually employing strong PUFs due to the infeasibility of fully characterizing all its CRPs in a reasonable time frame by an adversary~\cite{herder2014physical}. In the PUF enrolment phase, the server (\textit{verifier}) lodges a number of CRPs into its database. In the authentication phase when the PUF integrated device (\textit{prover}) is deployed in field, the server randomly picks up a challenge and sends it to the prover. Once the prover returns the measured response subject to the received challenge, the server compares the enrolled response in the database with the returned response from the prover. The authenticity of the prover is established only if the Hamming distance between the received and enrolled response is less than a predetermined threshold. However, due to various model building attacks~\cite{ruhrmair2013puf,becker2015gap,becker2015pitfalls,delvaux2017machine}, it is now recognized that it is hard to provide security guarantees for simple challenge-response based lightweight authentication protocols built upon silicon strong PUFs, specifically, variants of the APUF~\cite{gao2016obfuscated} and k-sum ROPUF~\cite{yu2014noise}. One method to halt modeling attacks is to limit the exposed number of CRPs by limiting the number of authentication rounds. The trade-off is that the PUF has to be disposed once a predetermined number of authentication rounds are reached~\cite{yulockdown}. 
A recent examination of strong PUF based authentication mechanisms  concluded that a secure PUF based authentication mechanism is better to be crafted from a PUF derived key~\cite{delvaux2017machine,delvaux2017physically}. \textit{Consequently, we will focus on key generation with PUFs and how secure key generation can be realized on resource limited devices}. A PUF based key derivation method has the potential to provide resource limited devices a cost effective authentication mechanism together with secure key storage. 
\vspace{1mm}

\noindent\textbf{Secure and Reliable Key Generation: }Unlike strong PUF responses that are usually correlated, weak PUF responses are normally independent from each other where each response is derived from an independent source of entropy. Consequently, a weak PUF, in general, is inherently immune to modeling attacks. Weak PUFs are highly suitable for deriving cryptographic keys. Recall that PUFs measure physical circuit properties to generate responses and---like any physical measurement---are inevitably affected by thermal noise and varying environmental conditions; hence, PUF response reproductions are not completely stable~\cite{hiller2016key}. Therefore, a PUF based key generator utilizes a fuzzy extractor to turn unstable responses into a stable and uniformly distributed cryptographic key. PUF based key generators have been well studied in the last decade, however, only small number of studies provide a complete implementation. Among the full implementations, majority are hardware implementations on FPGA platforms~\cite{van2012reverse,maes2012pufky} rather than software implementations on microcontroller (MCU) platforms~\cite{aysu2015end}. In fact, most IoT devices do not have FPGAs but operate using MCUs. Further, to the best of our knowledge, there is only one end-to-end software PUF key generator implementation on a low power computing platform in the form of a highly resource constrained and batteryless device that relies only on harvested RF energy for operations~\cite{Yang2018scode}.\textit{Therefore, we aim to provide a simple and timely guide to the Ubiquitous Computing community for realizing PUF based key generators to benefit from their inherent key security.} In particular:
\begin{itemize}
    \item We consider the use of freely available SRAM on low cost and low power MCUs to function as an intrinsic PUF to derive a secure and reliable cryptographic key.
    \item We consider a methodology for realizing lightweight implementations of reliable key generators capable of execution on low cost and ultra low power microcontrollers.
\end{itemize}

\section{Anatomy of a Key Generator}


\subsection{SRAM PUF}

SRAM memory is pervasively embedded within electronic commodities. When SRAM is powered up, each SRAM cell has a favored power-up state; see Fig.~\ref{fig:SRAMPUF}(b). Such a favored power-up state varies from cell to cell, and chip to chip. Therefore, the power-up pattern of SRAM memory can be treated as a PUF where the address of each cell is a challenge and power-up state the response. Thus SRAM yields an intrinsic PUF attributing to its wide scale availability on MCUs and its ability to be used without any overhead---extra hardware~\cite{holcomb2009power}; this makes SRAM PUFs a popular silicon PUFs nowadays. 

\subsection{(Reverse) Fuzzy Extractor}\label{sec:RFE}
\begin{figure}[h]
	\centering
	\includegraphics[trim=0 0 0 0,clip,width=0.30\textwidth]{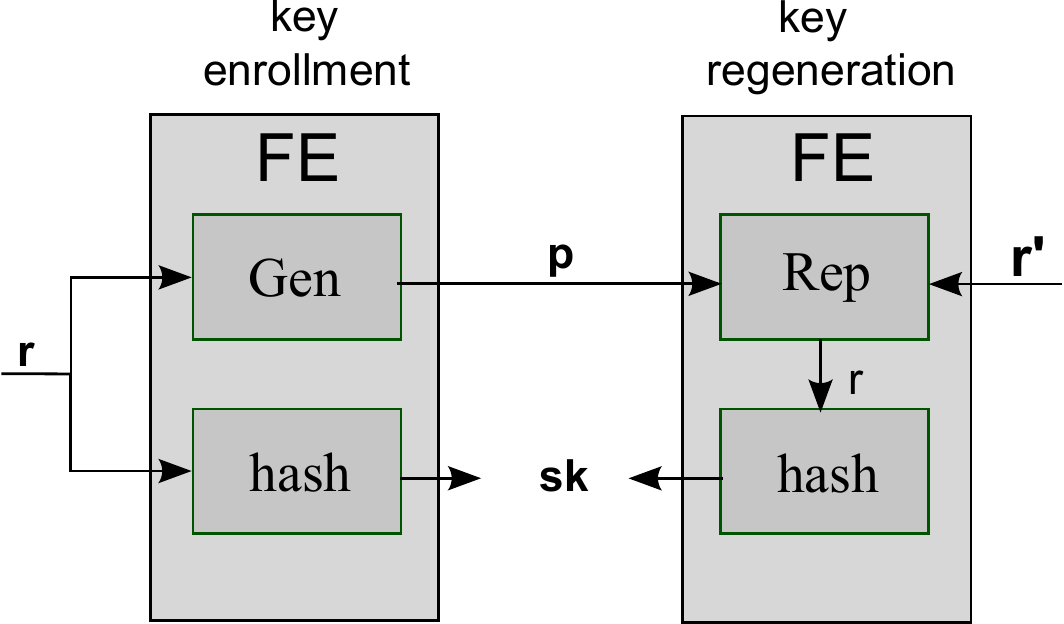}
	\caption{During the key enrollment phase, the \textsf{Gen}() takes enrolled response {\bf r} as input, and computes helper data {\bf p}. The enrolled secret key {\bf sk}=\textsf{hash}({\bf r}). During the key regeneration phase, the \textsf{Gen}() takes both the regenerated response ${\bf r}^{\prime}$ and the helper data {\bf p} as inputs, and recovers the enrolled response $\bf r$. The recovered $\bf r$ is hashed to obtain the enrolled secret key {\bf sk}.}
	\label{fig:keyGen}
\end{figure}

To turn the raw noisy PUF responses---for example, from an SRAM PUF---into a cryptographic key, a widely accepted approach is to employ a fuzzy extractor~\cite{maes2012pufky,delvaux2015helper,dodis2008fuzzy}---see the illustration in Fig.~\ref{fig:keyGen}. A fuzzy extractor consists of: i) a secure sketch; and ii) a procedure for entropy extraction. 


  In general, the secure sketch construction has a pair of functions: \textsf{Gen}() and \textsf{Rep}()---see overall illustration in Fig.~\ref{fig:keyGen}. There are two prevalent secure sketch constructions: i) code-offset construction; and ii) syndrome construction~\cite{delvaux2015helper}. We choose the later, mainly considering its security---see Section~\ref{sec:secDis}. During key enrollment phase, helper data $\bf p$ is computed by using \textsf{Gen}({\bf r}), where ${\bf p}={\bf r}\times {\bf H^T}$ and $\bf H$ is a parity check matrix of a linear error correction code. Key reconstruction described by \textsf{Rep}({\bf r}$^{\prime}$,{\bf p})---where ${\bf r}^{\prime}$ is the reproduced response that may be slightly different from the enrolled response $\bf r$---first constructs a syndrome ${\bf s}=({\bf r}^{\prime}\times{\bf H^T})\oplus {\bf p}={\bf e}\times{\bf H^T}$, with ${\bf e}$ an error vector. Then through an error location algorithm, $\bf e$ is determined. Subsequently, the response ${\bf r}$ is recovered through ${\bf r}={\bf e}\oplus{\bf r}^{\prime}$. The recovered PUF response $\bf r$ may not be uniformly distributed. Therefore, an entropy extraction step, using a universal hash function for example, is employed to generate a  cryptographic key with full bit entropy.
	
In a fuzzy extractor setting, the \textsf{Gen}() function is performed by the server during the enrollment phase to compute helper data while the \textsf{Rep}() function is implemented on the field deployed token. By recognizing that the computational burden of the \textsf{Rep}() function is significantly more than that of \textsf{Gen}() function, Van Herrewege {\it et al.}~\cite{van2012reverse} placed \textsf{Gen}() on the resource-constraint token  while leaving the computationally heavy \textsf{Gen}() function execution to the resource-rich server; this arrangement is termed the reverse fuzzy extractor (RFE).

The (R)FE must bear two properties: i)~{\bf Correctness}: It must be possible to correctly reconstruct {\bf r}---equivalently, the secret key $\bf sk$----given helper data $\bf p$ if the Hamming distance between the enrolled response {\bf r} and the reevaluated response ${\bf r}^{\prime}$ is less than an acceptable threshold; and ii)~{\bf Security}: Given exposed helper data $\bf p$, there must be adequate residual entropy in the PUF response $\bf r$ to ensure the security of the key $\bf sk$. 

Although the entropy extractor realized with a hash function can be lightweight, the overhead of the secure sketch is comparatively more dominant. \textit{This is a significant problem PUF based key generation on a resource constrained device, often limited by computational capability, memory, and power}. We will focus on the correct recovery in the context of resource limited devices and a RFE where the \textsf{Gen()} function is implemented on the resource limited token in Section~\ref{sec:RFE} and \ref{sec:Exp}. We discuss security in Section~\ref{sec:secDis}.



\section{Reducing Reverse Fuzzy Extractor Overhead}
\label{sec:RFE}

\subsection{Response Pre-Processing}
The overhead of implementing the \textsf{Gen()} function is directly related to the response unreliability or bit error rate (BER); noisier responses naturally require a higher error correction overhead. Therefore, it is imperative to reduce the expected BER of PUF responses on a token. In this context, response pre-processing methodologies, e.g., majority voting and reliable response pre-selection~\cite{bohm2013puf} can be employed. 

In majority voting (MV), given $q$ repeated response measurements under an operating condition, the response value (`1'/`0') enrolled is that which shows no less than $\ceil{\frac{q}{2}}$ measurements; with $q$ an odd integer. In reliable response pre-selection (PreSel), each PUF response under an operating condition is repeatedly measured $i$ times, only response bits that exhibit 100\% reliable regeneration (all `1's/`0's) are selected for use and enrollment---a detailed methodology with a resource limited device is in~\cite{Yang2018scode}.  
 
\subsection{Key Generation with Multiple Reference Responses}
Gao {\it et al.} recently proposed the multiple reference responses (MRR) enrollment strategy~\cite{gao2018lightweight}. This method is especially suitable in a reverse fuzzy extractor setting to significantly reduce the \textsf{Gen()} function implementation overhead on a token~\cite{gao2018lightweight}. The MRR rationale is described below in the context of a reverse fuzzy extractor, dubbed MR$^3$FE.

\begin{figure}[t]
	\centering
	\includegraphics[trim=0 0 0 0,clip,width=0.27\textwidth]{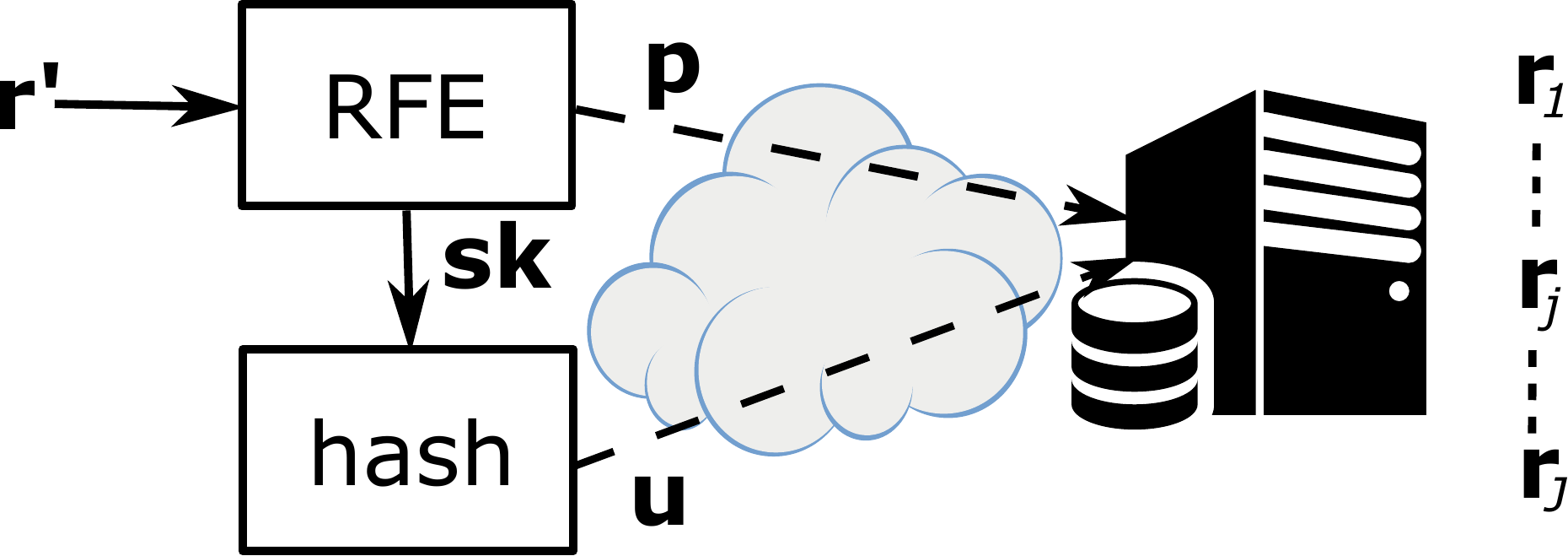}
	\caption{MR$^3$FE in the context of key generation on token.}
	\label{fig:RFEMRR}
\end{figure}

\begin{figure}[t]
	\centering
	\includegraphics[trim=1.8cm 0 0 0,clip,width=0.36\textwidth]{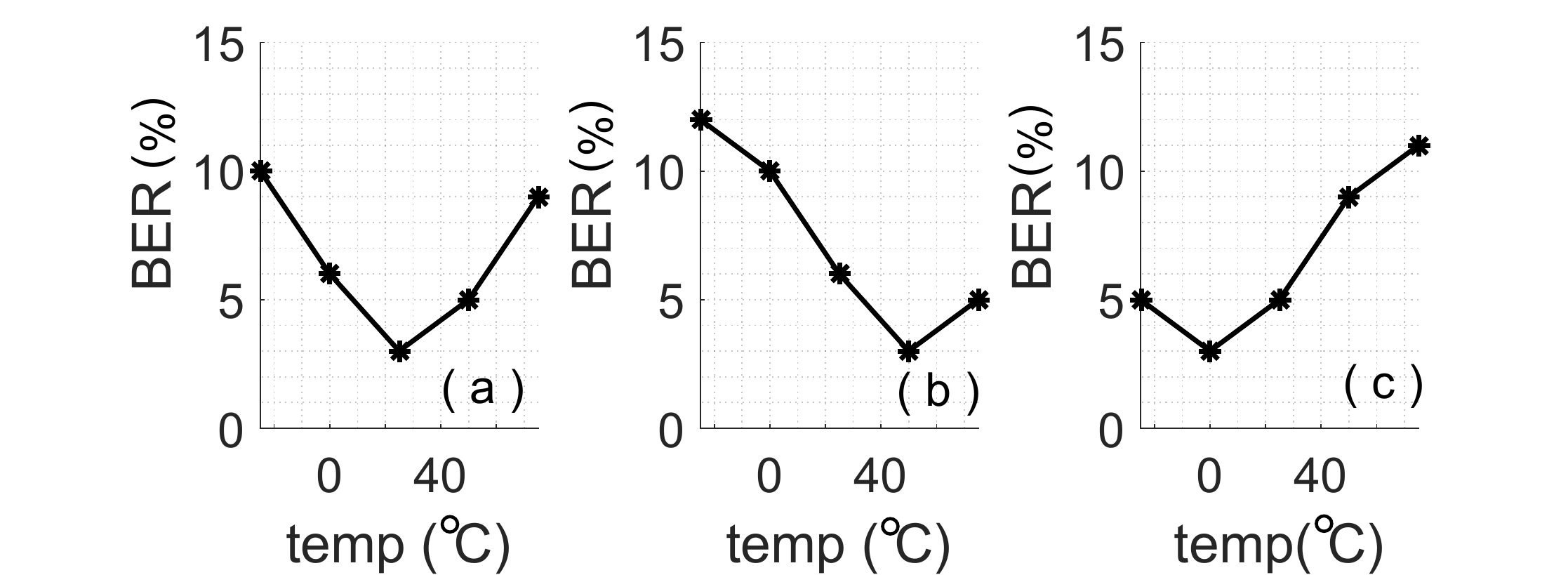}
	\caption{Bit error rate when the reference response is evaluated under: (a)~$25\celsius$; (b) $50\celsius$; and (c) $0\celsius$.}
	\label{fig:BERtwoOC}
\end{figure}

Fig.~\ref{fig:RFEMRR} illustrates MR$^3$FE in the context of key generation. During the enrollment phase, in contrast to conventional single response enrolment under a nominal operating condition, e.g., room temperature of $25\celsius$, the server enrolls $J$ reference responses $\{{\bf r}_1,...,{\bf r}_j,...{\bf r}_J\}$ subject to the same challenge $\bf c$ applied to the same PUF but under $J$ different operating conditions. We take two enrolled reference responses as an example. As shown in Fig.~\ref{fig:BERtwoOC}, we assume that the server enrolls two reference responses, ${\bf r}_1$ and ${\bf r}_2$, evaluated under $50\celsius$ and $0\celsius$, respectively.
We can observe in Fig.~\ref{fig:BERtwoOC}---regardless of the operating condition under which the reference response is enrolled---when the difference between the reference operating condition at enrollment increases with respect to the condition during a regenerated response, the the expected BER increases. 

When the PUF is deployed in the field, as shown in Fig.~\ref{fig:RFEMRR}, the token reevaluates ${\bf r}^{\prime}$ corresponding to a challenge $\bf c$, typically stored on chip---recall that we are using a reverse fuzzy extractor. Then the token computes helper data ${\bf p} \leftarrow $\textsf{Gen}$({\bf r}^{\prime})$ and generates secret key $\bf sk$ during run-time. The token sends $\bf p$ and ${\bf u} \leftarrow$\textsf{hash}(${\bf sk}||{\bf p}$) to the server---${\bf u}$ allows the server to check its success of recovering ${\bf sk}$, while also ensuring the integrity of the helper data $\bf p$ to alleviate potential helper data manipulation attacks~\cite{delvaux2015helper,becker2017robust}. The server now attempts to recover the response ${\bf r}^{\prime}$ based on two enrolled reference responses: ${\bf r}_1$ and ${\bf r}_2$. For example, the server first uses ${\bf r}_1$ to restore ${\bf r}^{\prime \prime} \leftarrow $\textsf{Rep}(${\bf r}_1$, $\bf p$). Then the server extracts a secret key ${\bf sk}^{\prime} \leftarrow$\textsf{hash}(${\bf r}^{\prime \prime}$) and consequently computes verification value ${\bf u}^{\prime}$=\textsf{hash}(${\bf sk}^{\prime}||{\bf p}$). If ${\bf u}^{\prime}={\bf u}$, then it implies that the ${\bf r}^{\prime \prime}$=${\bf r}^{\prime}$, equivalently, the $\bf sk$ is deemed to be correctly restored. Otherwise, $\bf sk$ regeneration fails based on ${\bf r}_1$. The server continues to use reference response ${\bf r}_2$ to try to regenerate $\bf sk$. 
Consequently, if any of the enrolled reference responses are capable of recovering $\bf sk$, MR$^3$FE is deemed successful and both the token and the server are in possession of the secret key $\bf sk$.
\vspace{1mm}

\noindent\textbf{MRR Advantages:}~We provide an intuitive illustration to explain. Let us firstly assume that we use the {\it conventional single reference response} enrolled under $25\celsius$ as shown in Fig.~\ref{fig:BERtwoOC}. We further assume that the regenerated response is evaluated under $-25\celsius$ where the expected BER is 10\% and suppose that the helper data $\bf p$ is only capable of correcting upto 6\% of response errors. It is clear that the RFE using reference response evaluated under $25\celsius$ to recover the secret key will most likely fail. Now given two reference responses ${\bf r}_1$ and ${\bf r}_2$ evaluated under $50\celsius$ and $0\celsius$, respectively. We can see that ${\bf r}_2$ is more likely to be able to recover the secret key $\bf sk$ even though the helper data $\bf p$ still has the same 6\% error correction capability because the expected BER at $-25\celsius$ using ${\bf r}_2$ as reference is only 5\%. What if the regenerated response ${\bf r}^{\prime}$ is evaluated under $80\celsius$? It is not difficult to see that now ${\bf r}_1$ is highly likely to successfully recover the secret key $\bf sk$. Although the server is unable to control the operating condition under which the response ${\bf r}^{\prime}$ is regenerated, we can observe that the server now has the capability of choosing one of multiple reference responses to ensure that the chosen reference response is close to the operating condition under which the ${\bf r}^{\prime}$ is regenerated. \textit{This ability greatly relaxes the error correction requirement and, thus, the implementation overhead of the \textsf{Gen()} function}.

\subsection{Key Failure Rate}
Our study uses the family of BCH($n$, $k$, $t$) linear codes with a syndrome based decoding strategy to realize a reverse fuzzy extractor considering its popularity~\cite{maes2012pufky,delvaux2015helper} and its security~\cite{delvaux2015helper,becker2017robust,gao2018lightweight}. Here, $n$ is the codeword length, $k$ is the code size, $t$ is the number of errors that can be corrected within this $n$-bit block. Assuming response bit errors are independently and identically distributed (i.i.d.), we can express the average key failure rate of recovering an $n$-bit  response ${\bf r}^{\prime} $ based on a selected reference response ${\bf r}_j$, termed as $\mathbb{P}_{1j}$, with $j\in \{1,..,J\}$ with $J$ the number of multiple references employed by the server, as:
\begin{equation}
\mathbb{P}_{1j}= 1 - \textsf{F}_B(t;n,{\rm BER}_j)
\end{equation}
where BER$_j$ is the expected BER using ${\bf r}_j$ as the reference response. Here, \textsf{F}$_B()$ is a cumulative density function of a binomial distribution with $t$ successes in $n$ Bernoulli trials, with each trial having success probability of BER$_j$.

A BCH($n$, $k$, $t$) encoding produces ($n-k$)-bit helper data that will be publicly known while the $k$ bits contribute to the secret. For a single BCH($n$, $k$, $t$) block, the complexity of finding the $k$-bit extracted key from the $n$-bit response $ {\bf r}^{\prime} $ is $2^{k}$. It is not common to use a single large BCH($n$, $k$, $t$) block; typically a large block is split into small processing blocks to reduce implementation complexity~\cite{hiller2016cherry}. For $k$ bits of key material, response $ {\bf r}^{\prime} $ can be divided into multiple non-overlapping blocks of a BCH($n_1$, $k_1$, $t_1$) code where $n_1<n$ and $k_1<k$ for a parallel implementation. Now the complexity of finding the $k$ bit secret is $2^{k_1\cdot L}$ where $L$ is the number of parallel BCH($n_1$, $k_1$, $t_1$) code blocks used to realize $k$ bits of secret key material. Given a BCH($n_1$, $k_1$, $t_1$) code employed to gain a security level of $k$ bits with $L=\ceil*{k/k_1}$ blocks, the key recovery failure rate under the assumption of i.i.d code blocks is:
\begin{equation}\label{eq:Pfail-Lblocks}
\mathbb{P}_{2j}= 1 - (1-\mathbb{P}_{1j})^L.
\end{equation}

When all $J$ reference responses $\{ {\bf r}_1,...,{\bf r}_j,...,{\bf r}_J \}$ are used, ${\bf r}^{\prime}$ reconstruction fails only when {\it all} reference responses cannot restore the response $ {\bf r}^{\prime} $. We adopt a very conservative evaluation of the key failure rate ${\mathbb{P}_{\rm fail}}$ expressed as\footnote{For a detailed discussion, the reader is referred to~\cite{gao2018lightweight}. }:
\begin{equation}
{\mathbb{P}_{\rm fail}}= min\{\mathbb{P}_{2j}\},~j\in\{1,...,J\}
\end{equation} 

\vspace{-0.2cm}
\section{Experimental Validations}\label{sec:Exp}
We first test the reliability of embedded SRAM PUFs. Then the software implementation set-up and the means of assessing clock cycle overhead is detailed. Finally, we implement the MR$^3$FE PUF key generator and validate the significant overhead reduction realized in comparison with: i) a single reference response based FE; and ii) a single reference response based RFE.
\vspace{-0.2cm}
\subsection{SRAM PUF Evaluation}
\subsubsection{SRAM PUF Dataset}
The PUF CRP dataset used is from 23 MSP430FR5969 microconrollers (MCUs) embedded in CRFID transponders. From each MCU, we read power-up states of 16,384 (2~KB) SRAM cells as SRAM PUF responses. The SRAM PUF reliability is much less sensitive to voltage variations compared with temperature fluctuations attributing to the SRAM cell's symmetric structure~\cite{selimis2011evaluation,roel2012physically}. Hence, we focus on its reliability under varying temperature conditions: $-15\celsius$, $0\celsius$, $25\celsius$, $40\celsius$ and $80\celsius$. Under each temperature condition, each response bit is repeatedly measured 100 times. 
\begin{figure} [t]
	\centering
	\includegraphics[trim=0cm 0 1cm 0.4cm,clip,width=0.25\textwidth]{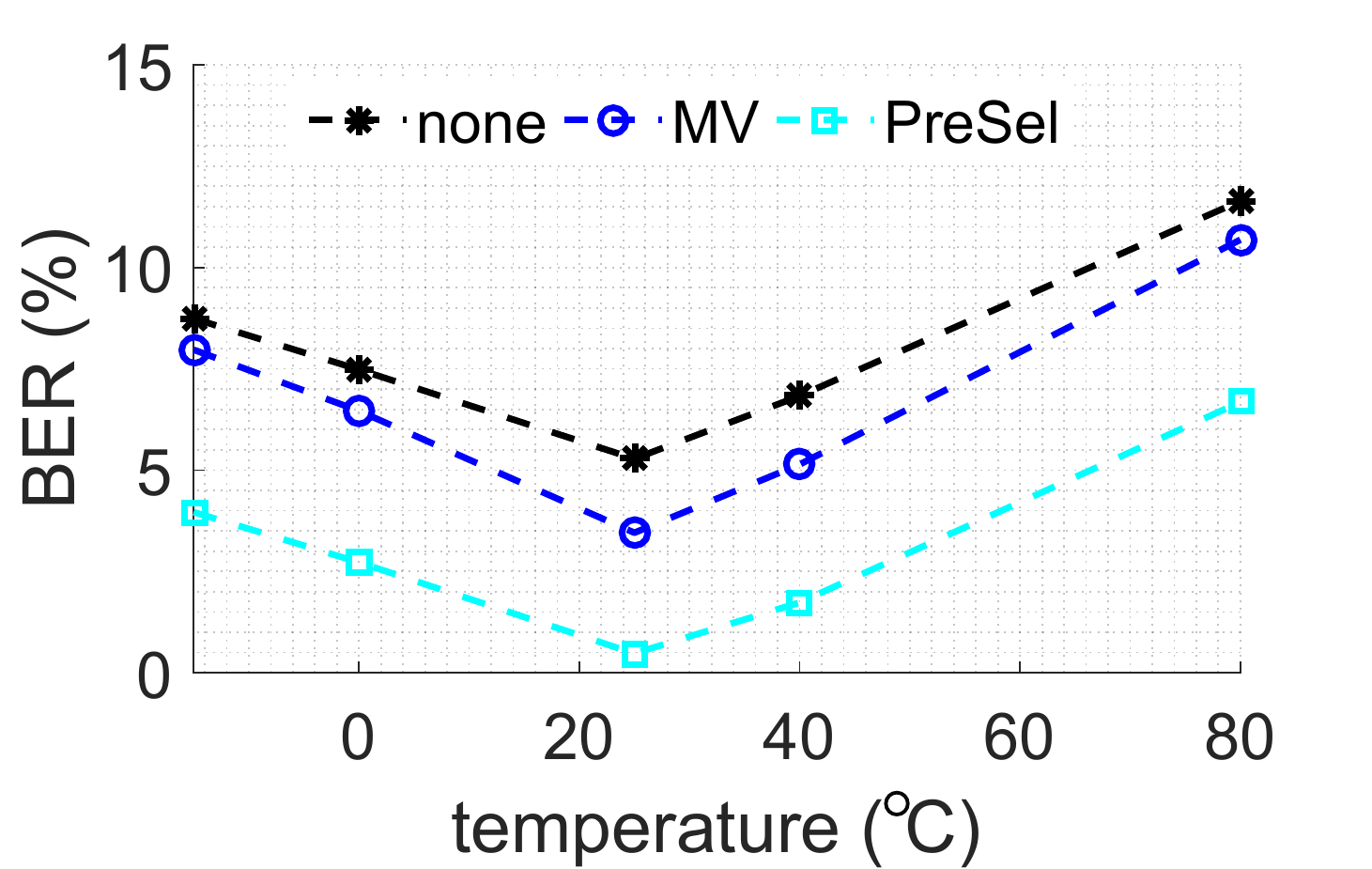}
	\caption{BER when reference response is enrolled under $25\celsius$.}
	\label{fig:BERMVPreSel}
\end{figure}

\subsubsection{Bit Error Rate}
As shown in Fig.~\ref{fig:BERMVPreSel}, for single reference response enrollment, BER under three different enrollment strategies are applied: one-time response measurement (none), majority voting (MV) with 9 repeated measurements ($q=9$), and preselection (PreSel) with 10 repeated measurements. The PreSel outperforms others.

\begin{figure} [t]
	\centering
	\includegraphics[trim=0cm 0 1cm 0,clip,width=0.45\textwidth]{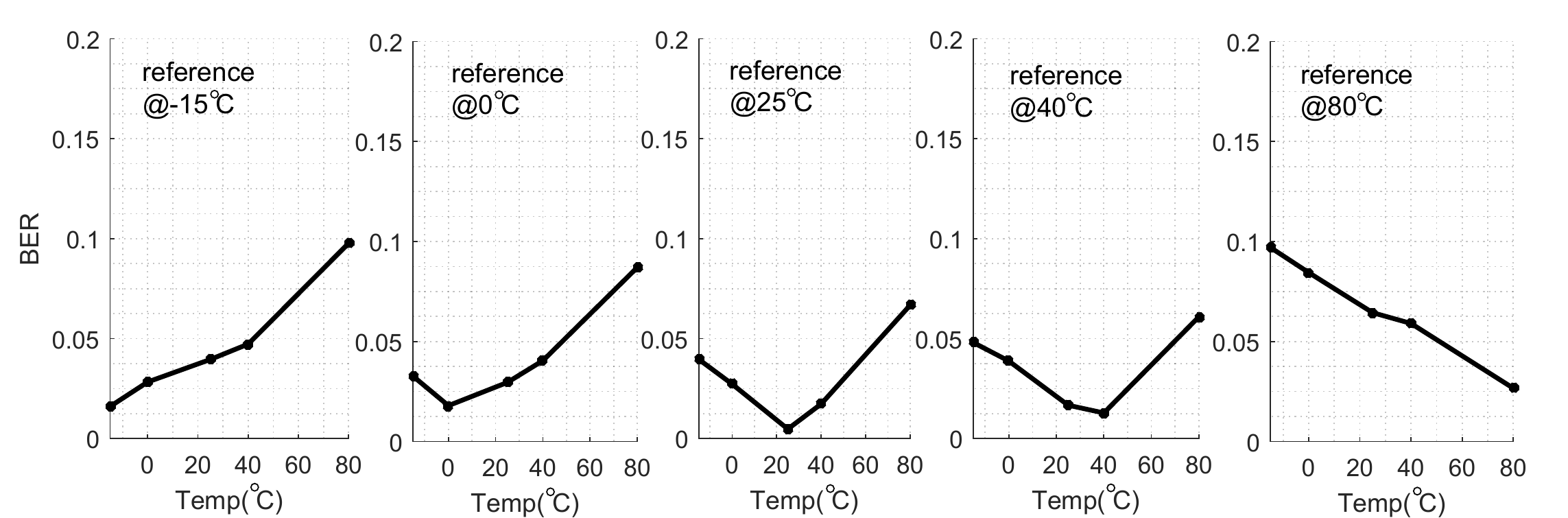}
	\caption{BER when a reference response is enrolled under  different operating conditions.}
	\label{fig:BER5OC_preselection}
\end{figure}

Fig.~\ref{fig:BER5OC_preselection} illustrates the BER when MRR are enrolled under differing operating conditions. For the reference response enrolled under $25\celsius$, preselection is applied to select reliable responses. Then reference response under  $-15\celsius$,  $0\celsius$,  $40\celsius$ and  $80\celsius$ applies response majority voting on the pre-selected reliable responses.
\vspace{-0.2cm}
\subsection{Overhead Evaluation}
\subsubsection{Testing Setup}
The test environment used is Texas Instruments' (TI) Code Composer Studio (CCS) 7.2.0, the C code used is downloaded to a MSP430FR5969 LaunchPad Evaluation Kit via USB. TI CCS has a built-in GCC toolchain for our hardware kit. This includes the \texttt{msp430-gcc-6.4.0.32} win32 compiler. The software instructions are executed sequentially as advanced out-of-order execution is unavailable for typical resource-constraint MCUs. The overhead of software implementation is assessed in terms of clock cycles to complete the algorithm---this is our primary evaluation measure. We measured clock cycles using Profile Clock tool supported in the CCS environment. 

Two key components of the PUF key generator are the hash function and the BCH code encoding/decoding--- \textsf{Gen()}/\textsf{Rep()}---blocks. Based on the hash function evaluations in~\cite{Yang2018scode}, we 
selected BLAKE2s-128 as it showed the best performance, in term of clock cycles. 
As for BCH($n_1, k_1, t_1$) code, it is imperative to use a small $n_1$ and $t_1$. First, the BCH($n_1, k_1, t_1$) code overhead decreases sharply when the codeword length $n_1$ decreases. Second, a small $t_1$ implies a correspondingly large $k_1$. With a large $k_1$, to achieve the security level of $k$-bit secret key $\bf sk$, the needed number of BCH($n_1, k_1, t_1$) blocks---$L=\ceil*{k/k_1}$---will reduce, thus, reducing the overall overhead to gain $k$-bit entropy for the secret key $\bf sk$. 

\subsubsection{Overhead Result and Comparison}
\begin{itemize}
\item \noindent{\bf Fuzzy Extractor with Single Reference Response:} To achieve $\mathbb{P}_{\rm fail} < 10^{-6}$, nine BCH(127,15,27) blocks are required giving $\mathbb{P}_{\rm fail} = 1.66\times 10^{-7}$ when the single reference response under $25\celsius$ is used. One BCH(127,15,27) block {\it decoding} consumes 2,102,222 clock cycles. The fuzzy extractor needs to sequentially execute BCH(127,15,27) decoding blocks 9 times and two BLACKE2s-128 hash operations. In total, it consumes up to 19,127,446 clock cycles.

\item \noindent{\bf Reverse Fuzzy Extractor with Single Reference Response:} To achieve $\mathbb{P}_{\rm fail} < 10^{-6}$, nine BCH(127,15,27) blocks are required giving $\mathbb{P}_{\rm fail} = 1.66\times 10^{-7}$ when the single reference response under $25\celsius$ is used. {\it Notably, the token only needs to perform the encoding operation requiring significantly less clock cycles in comparison with decoding}. One BCH(127,15,27) block {\it encoding} consumes 111,335 clock cycles. This reverse fuzzy extractor needs to sequentially execute BCH(127,15,27) encoding 9 times and two BLACKE2s-128 hash operations. In total, it consumes up to 1,211,461 clock cycles.

\item \noindent{\bf Reverse Fuzzy Extractor with Three Reference Response:} To achieve $\mathbb{P}_{\rm fail} < 10^{-6}$, eight smaller BCH(63,16,11) blocks are adequate to meet the requirement with $\mathbb{P}_{\rm fail} = 9.85\times 10^{-7}$. Here, 3MRR under $-15\celsius$, $25\celsius$, $80\celsius$ are used. One BCH(63,16,11) block {\it encoding} consumes 51,003 clock cycles. This reverse fuzzy extractor needs to sequentially execute BCH(63,16,11) encoding 8 times and two BLACKE2s-128 hash operations. In total, it consumes up to 617,470 clock cycles.
\end{itemize}

We can see that the reverse fuzzy extractor is especially suitable for resource limited devices attributing to the fact that the \textsf{Gen} overhead is significantly less than the \textsf{Rep} overhead. The reverse fuzzy extractor overhead is only 6.3\% of the fuzzy extractor when a single reference response is deployed to achieve the same key failure rate performance. When the MRR is adopted by the reverse fuzzy extractor, the clock cycle overhead is further reduced by 49\% in comparison with a single reference response. This is because smaller $n_1$ and $t_1$ code parameters---meaning less encoding overhead----of a BCH($n_1$,$k_1$,$t_1$) code can achieve the targeted key failure rate. It is worth mentioning here that the server has the ability to enroll more reference responses under fine-grained operating conditions to further reduce the MR$^3$FE overhead. Although this creates a \textit{one-time} burden on the server, the impact on the token through the reduction in overhead costs is significant and the benefits extend to the lifetime of the token in the field. 

\vspace{-0.2cm}
\section{Security Discussion}\label{sec:secDis}
Given a secure sketch with BCH($n$,$k$,$t$) code, entropy leakage is caused mainly from the public helper data. The well-known min-entropy loss is the $n-k$ bound given the exposure of helper data. This is mainly caused by the ($n-k$)-bit helper data exposure. Extra entropy leakage can be from response bias~\cite{maes2016secure,maes2016secure,delvaux2016efficient}; a bias exists when the probability of a response being `1' or `0' is not ideally 50\%.  Active helper data manipulation (HDM) attacks may also cause serious entropy leakage if the helper data algorithm is not carefully constructed~\cite{delvaux2014key,delvaux2014attacking,delvaux2015helper,becker2017robust}. More specifically, not all error correction codes and decoding strategies that are components of helper data algorithm for implementing the fuzzy extractor are able to guarantee the security of the PUF derived key~\cite{delvaux2015helper,becker2017robust}, therefore, when a helper data algorithm is introduced for realizing a fuzzy extractor, not only its implementation overhead but also its security against helper data manipulation attacks has to be carefully evaluated. Nonetheless, Our MR$^3$FE case study employed BCH codes and syndrome decoding, which has been shown to be secure under HDM attacks~\cite{becker2017robust}.

The reverse fuzzy extractor can result in unanticipated entropy loss under repeated helper data exposure associated with a given PUF response $\bf r^{\prime}$; unless, PUF responses are unbiased. Generally, the extra entropy loss is a result of the leakage of bit-specific reliability information~\cite{delvaux2016efficient}. However, the above extra entropy losses are important only when PUF response bias is considerably different from the ideal value of 50\% as shown by the analysis in~\cite{delvaux2016efficient,delvaux2017physically}. In practice, modern silicon PUFs usually have a good bias performance---close to 50\%~\cite{roel2012physically}. The bias of our tested SRAM PUF is 49.87\%. Therefore, the extra entropy loss is very small. In this context, employing a few more response bits in the reverse fuzzy extractor can compensate for the small extra loss in entropy. As observed by Delvaux~\cite{delvaux2017physically}, for a PUF with low bias within $[0.42, 0.58]$, increasing the length of raw responses alone is an effective measure.

If the bias is severe, entropy compensation by solely increasing the length of raw responses becomes ineffective. As a result, debiasing the biased raw responses~\cite{maes2016secure} must be undertaken first, e.g., via classic von Neumann (CVN) debiasing, pair-output von Neumann debiasing with erasures ($\epsilon$-2O-VN) and Hamming Weight (HW) based de-biasing~\cite{Yang2018scode}---an \textit{example} of employing a HW based debiasing method for key derivation with a resource limited devices is detailed in~\cite{Yang2018scode}. Notably, not all debiasing schemes offer reusability---multiple use of the same PUF response---for a reverse fuzzy extractor~\cite{delvaux2016efficient}. Thus, debiasing schemes, e.g., $\epsilon$-2O-VN, that offer reusability should be chosen for the reverse fuzzy extractor.

\vspace{-0.2cm}
\section{Conclusion}
We studied a reverse fuzzy extractor based PUF key derivation method with a focus on reducing the token implementation overhead. The overhead is reduced by reducing the response unreliability; realized by two compatible methodologies: i) reliable response preselection; and ii) multiple reference response enrollment performed during the PUF provisioning (enrollment) phase. For experiments, we used the intrinsic SRAM PUF---requires neither hardware modification nor extra area cost---to demonstrate a lightweight PUF key generator implementation on a battery-less CRFID device, followed by security analyses. 
\section{Acknowledgement}
We acknowledge support from Australian Research Council Discovery Program (DP140103448), NJUST Research Start-Up Funding (AE89991/039, AE89991/103) and the National Natural Science Foundation of China (61802186).
\vspace{-0.2cm}

\end{document}